\documentclass[11pt,a4paper,superscriptaddress,aps,preprint]{revtex4}
\usepackage{amsfonts}
\usepackage{graphics}
\usepackage{graphicx}
\usepackage{epsf}
\begin{document}

\title{Aharonov-Bohm Effect on Noncommutative Plane: A Coherent State Approach}
\author{M. A. Anacleto, J. R. Nascimento, A. Yu. Petrov}
\affiliation{Departamento de F\'{\i}sica, Universidade Federal da Para\'{\i}ba\\
 Caixa Postal 5008, 58051-970, Jo\~ao Pessoa, Para\'{\i}ba, Brazil}
\email{anacleto,jroberto,petrov@fisica.ufpb.br}


\begin{abstract}
We formulate in a systematic manner the coherent state approach and apply it to study Aharonov-Bohm effect in the field theory context. Within this approach, we verify that the scattering amplitude is ultraviolet finite. Also, we prove that introduction of a quartic self-interaction for the scalar field allows to obtain a smooth commutative limit. 
\end{abstract}

\maketitle

\section{Introduction}
In the last years, noncommutative theories have been discussed in
the literature by a large number of authors, mainly due to the
discovery of their relation to string theory \cite{Seiberg}.
Noncommutative field theories were obtained by replacing the usual
product of fields by the Moyal product (star product). Because of
the properties of star product, the quadratic part in the action is
the same as in the commutative case, with only interaction terms are
modified. Consequently, the free propagator is not changed after introducting of the
star product. The inherent nonlocality of these theories leads to
the surprising mixture between ultraviolet (UV) and infrared (IR)
divergences which could break the perturbative expansion. Besides, the Feynman diagrams turn out to exhibit the same (up to the numerical coefficients)
ultraviolet divergences as in the commutative case.

To avoid difficulties with the UV/IR mixing, recently, a new approach was developed in \cite{Spallucci, Spallucci2} to study the noncommutative
space-time. This new formalism is based on the coherent state
approach, instead of Moyal product approach, and is explicitly
ultraviolet finite. In this approach the free propagators
acquire a Gaussian  damping factor, which incorporates $\theta$
as a natural cutoff at large momenta. In \cite{Jahan} this
formalism was applied for the nonrelativistic scalar field theory.
It was shown that this theory is ultraviolet finite on a
quantum plane at the one-loop level. In particular, the divergent
behavior of the model has been regained in the limit
$\theta\rightarrow 0$, i.e., when the noncommutativity of
the coordinates is removed.

In this work we will apply the coherent states approach to the
Aharonov-Bohm (AB) effect in the noncommutative plane. In quantum
field theory, this effect represents itself essentially as the scattering of spin
zero particles through a Chern-Simons (CS) field. It is well-known
that, in the commutative situation \cite{Lozano}, the quartic
self-interaction is necessary to ensure the renormalizability of
the model. In the context of the Moyal product based noncommutative nonrelativistic field
theory \cite{Anacleto}, it has been shown that up to the one-loop
order the UV divergences of the planar contributions are canceled
for the four-point function. On the other hand,
to remove logarithmic infrared divergences originating from the
non-planar diagrams in the commutative limit, it was necessary to
introduce  a quartic self-interaction to the Lagrangian.

We show that up to the one-loop order the scattering amplitude is ultraviolet finite.
At the same time, the amplitude involves logarithmic singularities as the noncommutative 
parameter tends to zero. 
To eliminate them, it is necessary to include a quartic self-interaction for the scalar 
field.

The paper is organized as follows. In Sec. II, we give a brief review of the coherent state approach to the noncommutative plane.
In Sec. III, we introduce the model and we compute the particle-particle scattering up to order one loop.
Final comments are made in the Sec. IV.

\section{Coherent State Approach}

In this section we present a brief description of the coherent state approach suggested in \cite{Spallucci,Spallucci2}.
We start with noncommutative plane described by the coordinates $\hat{q}_{1}$ and $\hat{q}_{2}$ satisfying
\begin{equation}
[{\hat{q}}_{1}, \hat{q}_{2}]=i\theta,
\end{equation}
where $\theta$ is the noncommutative parameter.
Let us now introduce a set of operators defined as
\begin{eqnarray}
\hat{A}&\equiv&\hat{q}_{1}+i\hat{q}_{2},
\nonumber\\
\hat{A}^{\dagger}&\equiv&\hat{q}_{1}-i\hat{q}_{2}.
\end{eqnarray}
The above operators satisfy commutation relation
\begin{equation}
[\hat{A}, \hat{A}^{\dagger}]=2\theta.
\end{equation}

Coherent states corresponding to the new operators are introduced as the eigenstates
$\left|\alpha \right\rangle$ in the following sense
\begin{eqnarray}
\hat{A}\left|\alpha \right\rangle=\alpha\left|\alpha \right\rangle,
\nonumber\\
\left\langle\alpha\right|\hat{A}^{\dagger}=\left\langle\alpha\right|\alpha^{\ast},
\end{eqnarray}
with normalized coherent states, $\left\langle\alpha\left.|\right.\alpha \right\rangle =1$, are defined as
\begin{equation}
\left|\alpha \right\rangle={\mbox{exp}}\left(-\frac{\alpha\alpha^{\ast}}{2}\right){\mbox{exp}}\left(-\alpha \hat{A}^{\dagger}\right)\left|0 \right\rangle,
\end{equation}
where the vacuum state $\left|0 \right\rangle$ is annihilated by $\hat{A}$.

The mean position of the particle over the noncommutative plane is defined as
\begin{eqnarray}
x_{1}&\equiv&\left\langle\alpha\right|\hat{q}_{1}\left|\alpha \right\rangle,
\nonumber\\
x_{2}&\equiv&\left\langle\alpha\right|\hat{q}_{2}\left|\alpha \right\rangle.
\end{eqnarray}
Thus, to any operator $F(\hat{q}_{1},\hat{q}_{2})$ we associate an ordinary function $f(x_{1},x_{2})$
through their mean values as follows:
\begin{eqnarray}
\label{func}
f(x_{1},x_{2})&=&\left\langle\alpha\right|F(\hat{q}_{1},\hat{q}_{2})\left|\alpha \right\rangle
\nonumber\\
&=&\int\frac{d^{2}k}{(2\pi)^{2}}f(k)\left\langle\alpha\right|e^{i\hat{\bf q}\cdot{\bf p}}\left|\alpha \right\rangle
\nonumber\\
&=&\int\frac{d^{2}k}{(2\pi)^{2}}f(k)e^{-\frac{\theta}{2}k^{2}}e^{i{\bf x}\cdot{\bf p}}.
\end{eqnarray}

Mean values of any operator over coherent states are commutative
quantities upon which one can construct usual quantum field
theory.  
It is very natural to expect that presence of the
exponential factors in the propagator would imply in the UV
finiteness of the corresponding field theory \cite{Madore}.

\section{Noncommutative Perturbative Theory}
Now let us apply the coherent state approach to the model of a nonrelativistic scalar field coupled
with a Chern-Simons field in 2+1 dimensions characterized by the
action
\begin{eqnarray}
S[A,\phi]\!\!\!&=&\!\!\!\!\! \int \!\!d^3x \left\{\frac{\kappa}{2} \epsilon^{ij}(A_0 F_{ij}-A_i\partial_t A_j)
- \frac{1}{2\xi}\partial_{i}A^{i}\partial_{j}A^{j}
+i\phi^{\dagger} D_{t}\phi -\frac{1}{2m}
({\bf{D}}\phi)^{\dagger}({\bf{D}}\phi)-\frac{\lambda_{0}}{4}
\phi^{\dagger}\phi^{\dagger}\phi\phi \right\}\!.
\label{accao}\nonumber\\
\end{eqnarray}
where the covariant derivatives are given by
\begin{eqnarray}
D_{t}\phi &=&\partial_{t}\phi+igA_{0} \phi,  \nonumber\\
D_{i}\phi &=&\partial _{i}\phi+igA_{i} \phi.
\end{eqnarray}
This action is commutative, i.e., the product of fields in
(\ref{accao}) is not a Moyal  product, but an ordinary product of
functions. The noncommutativity will be implemented in this model by
modification of the propagators via incorporation of the Gaussian
factors that appear in the Eq. (\ref{func}).

\begin{figure}
\centering
{\includegraphics{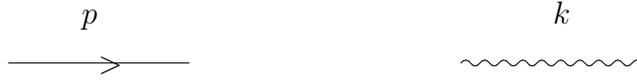}}
\caption{Propagators.}
\label{propagadores}
\end{figure}

\begin{center}
\begin{figure}
\centering
{\includegraphics{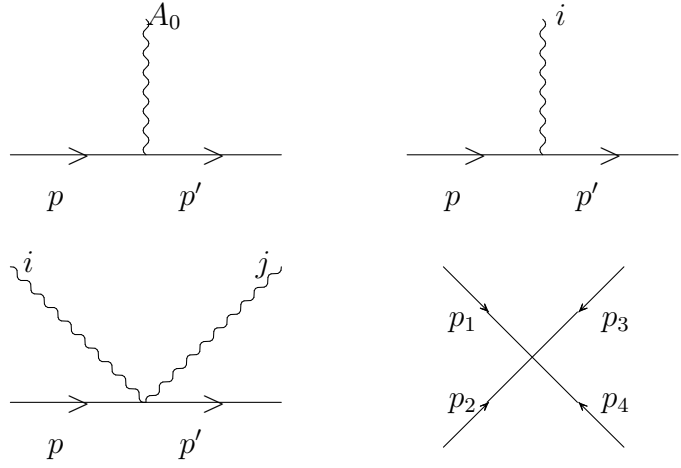}}
\caption{Vertices.}
\label{vertices}
\end{figure}
\end{center}

For convenience, we will work in the Coulomb gauge by choosing $\xi\rightarrow0$. 
Furthermore, we will use a graphical notation where the CS field and the matter field propagators are represented by wavy and continuous lines, respectively (see Fig. \ref{propagadores}).
The analytical expressions for the matter and gauge field propagators  are now given by
\begin{eqnarray}
D(p)&=&\frac{ie^{-\theta {\bf p}^2/2}}{p_{0}-\frac{\bf{p}^2}{2m} + i\epsilon},
\\
D_{i0}(k)&=&-D_{0i}(k)=\frac{\varepsilon_{ij}{\bf k}^{j}}{\kappa {\bf k}^{2}}e^{-\theta {\bf k}^2/2}
\end{eqnarray}
The analytical expressions associated with the interactions vertices (see Fig. \ref{vertices}) are
\begin{eqnarray}
&&\Gamma^{0}=-ig, \\
&&\Gamma^{i}=\frac{ig}{2m}{(\bf{p}+\bf{p}^{\prime})}^{i}, \\
&&\Gamma^{ij}=-\frac{ig^2}{m}\delta ^{ij}, \\
&&\Gamma=-i\lambda_{0}.
\end{eqnarray}
Due to the momentum conservation, these vertices are not affected by the noncommutativity in the coherent state formalism. This is a very important statement. We illustrate it taking the $\Gamma^{0}$ vertex as an example:
\begin{equation}
\label{ver}
-g\int d^3x\phi^{\dagger}A_{0}\phi=-g\int d^3x
\int d^3kd^3pd^3q<\alpha|e^{i(k-p-q)x}|\alpha>\tilde{\phi}(k)\tilde{\phi}(p)\tilde{\phi}(q).
\end{equation}
The phase factor is evaluated as
\begin{eqnarray}
<\alpha|e^{i(k-p-q)_{j}x^{j}}|\alpha>&=&<\alpha|e^{i(k-p-q)_{1}x_{1}+ i(k-p-q)_{2}x_{2}}|\alpha>
\nonumber\\
&=&<\alpha|e^{i(k-p-q)_{+}\hat{A}^{\dagger}+ i(k-p-q)_{-}\hat{A}}|\alpha>
\nonumber\\
&=&<\alpha|e^{i(k-p-q)_{+}\hat{A}^{\dagger}}\quad e^{i(k-p-q)_{-}\hat{A}}\quad e^{(k-p-q)_{+}(k-p-q)_{-}[\hat{A}^{\dagger},\hat{A}]/2}|\alpha>
\nonumber\\
&=&e^{i({\bf k}-{\bf p}-{\bf q})\cdot{\bf x}}\quad e^{-\frac{\theta}{2}({\bf k}-{\bf p}-{\bf q})^{2}},
\end{eqnarray}
where we have considered  $$(k-p-q)_{\pm}=\frac{(k-p-q)_{1}\pm i(k-p-q)_{2}}{\sqrt{2}}.$$
So that the Eq. (\ref{ver}) takes the form
\begin{eqnarray}
&-&g\int d^3x\phi^{\dagger}A_{0}\phi=-\frac{g}{(2\pi)^{6}}\int d^{3}kd^{3}pd^{3}q\quad e^{-\frac{\theta}{2}({\bf k}-{\bf p}-{\bf q})^{2}}\delta^{3}(k-p-q)\tilde{\phi}(k)\tilde{\phi}(p)\tilde{\phi}(q)\nonumber\\&=&-\frac{g}{(2\pi)^{6}}\int d^{3}k d^{3}p d^{3}q\quad \delta^{3}(k-p-q)\tilde{\phi}(k)\tilde{\phi}(p)\tilde{\phi}(q),
\end{eqnarray}
which coincides with the commutative analog of this vertex. Thus we conclude that the vertices are not modified within the coherent state approach. It is easy to show that the same situation occurs for other vertices, cf. the Refs \cite{Spall, Jahan}. Now we are ready to formulate a general prescription for calculating the loop contributions within this approach which consists in modifying the propagators by introduction  the Gaussian phase factors whereas the vertices are not modified.    

\begin{center}
\begin{figure}
\centering
{\includegraphics{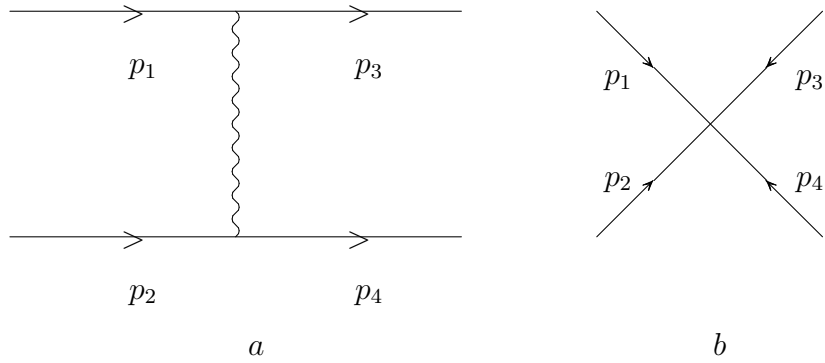}}
\caption{Tree-level scattering.}
\label{treelevel}
\end{figure}
\end{center}

Let us start our analysis by evaluating the four-point function associated with the scattering of two identical particles in the center-of-mass frame.
In the tree approximation the gauge part of the two body scattering amplitude is presented graphically in Fig. \ref{treelevel}(a) corresponding to the following analytical expression:
\begin{equation}
{\cal{A}}_{a}^{0}(\varphi)=-\frac{ig^2}{m\kappa}
\left[\frac{e^{-\bar{\theta}(1-\cos\varphi)}}
{1-\cos\varphi}
-\frac{e^{-\bar{\theta}(1+\cos\varphi)}}
{1+\cos\varphi}\right]\sin\varphi. \label{tree}
\end{equation}
Here we introduce a new dimensionless noncommutativity parameter $\bar\theta =\theta {\bf p}^{2}$. 
The second term in (\ref{tree}) corresponds to the crossed graph, where the final particle states are exchanged.  
The amplitude (\ref{tree}) for small $\bar{\theta}$  takes the form
\begin{equation}
{\cal{A}}_{a}^{0}(\varphi)=-\frac{2ig^2}{m\kappa}\left(\cot\varphi-\frac{\bar{\theta}^2}{4}\sin 2\varphi\right),
\end{equation}
where $\varphi$ is scattering angle between the incoming (${\bf p}$) and the outgoing (${\bf p}^{\prime}$) momenta. Here and further the terms of higher orders in $\theta$ are omitted.
We would remind that contributions near $\varphi=0$ are not well defined for
the Aharonov-Bohm scattering \cite{Jackiw}. 
The first noncommutative correction occurs only in second order in the parameter $\bar{\theta}$ and contains an angular dependence.  
The tree-level scattering amplitude obtained by means of the use of a Moyal product \cite{Anacleto} is given by
\begin{equation}
{\cal{A}}_{a}^{0}(\varphi)=-\frac{ig^2}{m\kappa}
\left[\frac{e^{i\bar{\theta}\sin\varphi}}
{1-\cos\varphi}
-\frac{e^{-i\bar{\theta}\sin\varphi}}
{1+\cos\varphi}\right]\sin\varphi,
\end{equation}
whose expansion at small values of $\bar{\theta}$ looks like
\begin{equation}
{\cal A}_{a}^0(\varphi)=-\frac{2ig^2}{m\kappa}\left(\cot\varphi
+i\bar{\theta}-\frac{\bar{\theta}^2}{4}\sin 2\varphi\right).
\end{equation}
We observe that due to different forms of introduction of noncommutativity the first-order correction in $\bar{\theta}$ is not generated within the coherent state approach while the second order corrections coincide within  both approachs.

By taking into account the quartic self-interaction, shown in Fig. \ref{treelevel}(b), the full tree level amplitude within the coherent state approach is
\begin{equation}
{\cal{A}}^{0}(\varphi)=-\frac{2ig^2}{m\kappa}\left(\cot\varphi-\frac{\bar{\theta}^2}{4}\sin 2\varphi\right)-\lambda_{0}.
\end{equation}

\begin{figure}
\centering
{\includegraphics{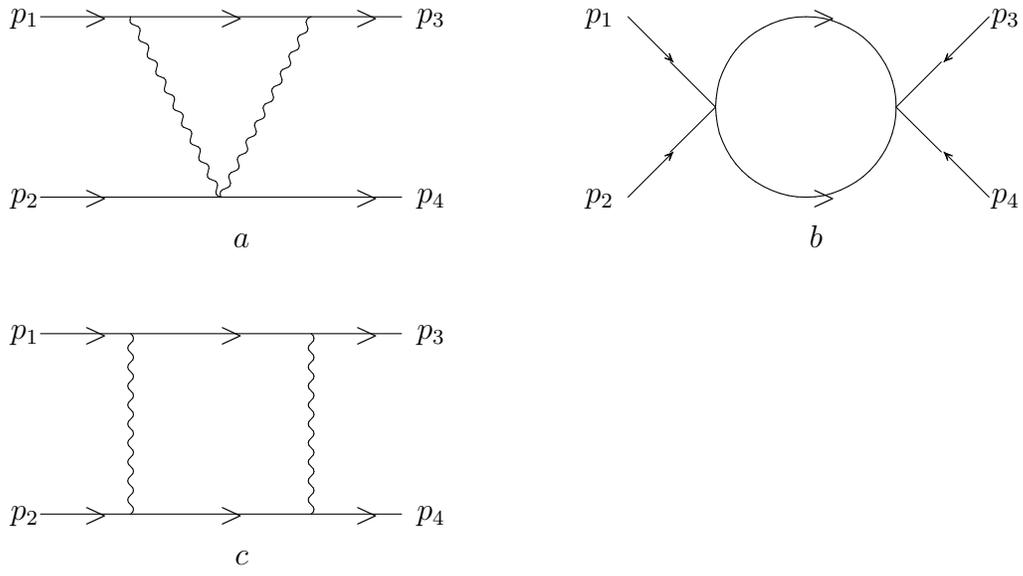}}
\caption{One-loop contributions.}
\label{loop}
\end{figure}

Let us now calculate the scattering at one loop order. The relevant diagrams are depicted in Fig. \ref{loop} 
(all other possible one-loop graphs vanish).
The contribution for the triangle graph drawn in Fig. \ref{loop}(a) after performing the $k_{0}$ integration is
\begin{eqnarray}
\label{tri}
{\cal{A}}_{a}(\varphi)&=&-\frac{g^{4}}{m\kappa^{2}}
\int\frac{d^{2}{\bf k}}{(2\pi)^{2}}\frac{{\bf k}\cdot({\bf k}-{\bf q)}}
{{\bf k}^{2}({\bf k}-{\bf q})^{2}} e^{-\theta[{\bf k}^{2}+({\bf k}-{\bf q})^{2}+({\bf p}_{1}-{\bf k})^{2}]/2}
+({\bf p}_{3}\leftrightarrow -{\bf p}_{3}) ,
\end{eqnarray}
where ${\bf q}={\bf p}_{1}-{\bf p}_{3}$ is the momentum
transferred. This integral can be evaluated analytically using the
Feynman parametrization, and the result is
\begin{eqnarray}
{\cal{A}}_{a}(\varphi)&=&-\frac{g^{4}}{m\kappa^2}
\int^{1}_{0} dx\int\frac{d^2{\bf k}}{(2\pi)^{2}}\frac{{\bf{k}}^{2}-a^{2}+(1-2x)\bf{q}\cdot\bf{k}}
{({\bf{k}}^{2}+a^{2})^{2}} e^{-\frac{3\theta}{2}[({\bf k}-{\bf q}x-{\bf p}_{3}-{\bf Q}/3)^{2}]}
\nonumber\\
&&+({\bf p}_{3}\leftrightarrow -{\bf p}_{3}),
\end{eqnarray}
where $a^{2}={\bf q}^{2}x(1-x)$ and ${\bf Q}={\bf p}_{1}+{\bf p}_{3}$.
Now using  standard Schwinger parametrization and performing
the $\bf k$ integration we have
\begin{eqnarray}
{\cal{A}}_{a}(\varphi)&=&-\frac{g^{4}}{4\pi m\kappa^{2}}
\int^{1}_{0} dx\int^{\infty}_{0}ds\left[ \frac{1-2a^2s}
{(s+\frac{3\theta}{2})}
+\frac{3s}{(s+\frac{3\theta}{2})^{2}}\left(\theta{{\bf q}}^{2}x(1-x)-\frac{\theta {\bf q}^2}{4}\right)\right]
e^{-sa^{2}}
\nonumber\\
&&{\mbox{exp}}\left[\frac{3\theta}{2}[{\bf Q}^2+{\bf q}^2x(1-x)]-\frac{5}{2}\theta {{\bf p}}^2\right]
+({\bf p}_{3}\leftrightarrow -{\bf p}_{3}).
\end{eqnarray}
Using the result
\begin{equation}
E_{i}(\alpha x)=\int^{\infty}_{x}dt\frac{e^{-\alpha t}}{t}=-\gamma-\ln(\alpha x)-\sum^{\infty}_{n=1}\frac{(-1)^{n}(\alpha x)^{n}}{nn!},
\end{equation}
the symbol $\gamma$ is the Euler-Mascheroni constant.
For small $\bar \theta$ we obtain
\begin{equation}
{\cal{A}}_{a}(\varphi)=\frac{g^{4}}{2\pi m\kappa^{2}}
\left[\gamma\left(1-\frac{3\bar{\theta}}{4}\right)+\ln(3/2)+\ln\bar{\theta}+\ln(2\sin\varphi)-\frac{25\bar{\theta}}{6}-\frac{\bar{\theta}}{2}\ln(2\sin\varphi)\right].
\end{equation}

The bubble graph shown in Fig. \ref{loop}(b) after performing $k_{0}$ integration is given by
\begin{eqnarray}
\label{cont}
{\cal{A}}_{b}(\varphi)&=&\frac{m\lambda_{0}^{2}}{8\pi}
\int_{0}^{\infty}d({\bf k}^{2})\frac{e^{-\theta {\bf{k}}^{2}}}{{{\bf(k}}^{2}-{\bf{p}}^{2}
-i\epsilon)}.
\end{eqnarray}
Using the well-known decomposition
\begin{equation}
\frac{1}{{\bf{k}}^{2}-{\bf {p}}^{2}-i\epsilon}=P\frac{1}{{\bf{k}}^{2}-{\bf{p}}^{2}}+i\pi\delta({\bf{k}}^{2}-{\bf{p}}^{2}),
\end{equation}
we can split the amplitude into a real part and an imaginary part. The real part yields
\begin{equation}
Re[{\cal{A}}_{b}(\varphi)]=-\frac{m\lambda_{0}^{2}}{8\pi}e^{-\bar{\theta}}\left[\gamma + \ln\bar{\theta} + \sum_{n=1}^{\infty}\frac{\bar{\theta}^{n}}{nn!}\right],
\end{equation}
and the imaginary part is
\begin{equation}
Im[{\cal{A}}_{b}(\varphi)]=i\frac{m\lambda_{0}^{2}}{8}e^{-\bar{\theta}}.
\end{equation}
For small $\bar{\theta}$ the amplitude becomes
\begin{eqnarray}
{\cal{A}}_{b}(\varphi)&=&-\frac{m\lambda_{0}^{2}}{8\pi}[\gamma(1-\bar{\theta}) + \ln\bar{\theta}+\bar{\theta}-i\pi(1-\bar{\theta})].
\end{eqnarray}

The contribution of the box diagram in Fig. \ref{loop}(c) after performing $k_{0}$ integration is
\begin{equation}
\label{box}
{\cal{A}}_{c}(\varphi)=\frac{4g^{4}}{m\kappa^{2}}
\int \frac{d^{2}{\bf k}}{(2\pi)^{2}}\frac{({\bf{p}}_{1}\wedge
{\bf{k}})\cdot({\bf{p}}_{3}\wedge
{\bf{k)}}e^{-\theta[({\bf{k}}-{\bf{p}}_{1})^{2}+ ({\bf{k}}-{\bf{p}}_{3})^{2} + 2{\bf{k}}^{2}]/2}}
{({\bf{k}}-{\bf{p}}_{1})^{2}
({\bf{k}}-{\bf{p}}_{3})^{2}({\bf{k}}^{2}-{\bf{p}}^{2}-i\epsilon)}
+ ({\bf p}_{3}\leftrightarrow -{\bf p}_{3}),
\end{equation}
where ${\bf k}\wedge{\bf p}$ is a vector product of the corresponding vectors.
This integral is finite, and for small $\theta$ we have
\begin{equation}
{\cal{A}}_{c}(\varphi)=-\frac{g^{4}}{2\pi m\kappa^{2}}[\ln(2\sin\varphi)+i\pi] 
+ \frac{\bar{\theta}g^4}{2\pi m\kappa^2}\left[2 + \ln(2\sin\varphi) + 2\cos\varphi\ln\left(\tan\left(\frac{\varphi}{2}\right)\right) + i\pi \right].
\end{equation}
Thus, summing all the results, we get
\begin{eqnarray}
{\cal{A}}(\varphi)&=&{\cal{A}}_{a}(\varphi)+{\cal{A}}_{b}(\varphi)+{\cal{A}}_{c}(\varphi)
\nonumber\\
&=&\frac{1}{4\pi m}\left(\frac{4g^{4}}{\kappa^{2}}-m^{2}\lambda_{0}^{2}\right)[\gamma + \ln\bar{\theta}-i\pi] + \frac{g^{4}}{2\pi m\kappa^2}\ln(3/2)-\frac{\bar{\theta}m\lambda_{0}^{2}}{8\pi}(1-\gamma + i\pi)
\nonumber\\
&&+ \frac{\bar{\theta}g^4}{2\pi m\kappa^2}\left[\frac{1}{2}\ln(2\sin\varphi) + 2\cos\varphi\ln\left(\tan\left(\frac{\varphi}{2}\right)\right) -\frac{13}{3}-\frac{3}{4}\gamma + i\pi \right].
\end{eqnarray}
This amplitude displays a logarithmic singularity at
$\bar{\theta}=0$. 
Arising of such a
singularity in the commutative limit is a natural consequence of
introduction of noncommutativity which plays the role of the UV
regulator of the theory. We note, however, that such a regulator is not a matter of choice but emerges naturally from the coherent states formalism. Effectively arising of the singularities at $\bar{\theta}=0$ shows that the theory has the correct commutative limit. 
Notice that the result for the commutative analog of this theory \cite{Lozano}, with $\Lambda^2$ is an
ultraviolet cutoff $(\Lambda \rightarrow\infty)$, is reproduced for
$\theta=1/\Lambda^{2}$.
The renormalization of this amplitude is
implemented by redefining the nonrelativistic self-coupling
constant $\lambda_{0}$:
\begin{eqnarray}
\lambda_{0}&=&\lambda + \delta\lambda,
\nonumber\\
\delta\lambda&=&\frac{1}{2\pi m}\left(\frac{4g^{4}}{\kappa^{2}}-m^{2}\lambda^{2}\right)
\ln(\theta\mu^{2})+\frac{g^{4}}{2\pi m\kappa^{2}}\ln(3/2) + {\cal O}(\lambda^{3},g^{6}),
\end{eqnarray}
and the total renormalized amplitude is given by
\begin{eqnarray}
{\cal{A}}(\varphi)&=&-\frac{2ig^2}{m\kappa}\cot\varphi-\lambda + \frac{1}{4\pi m}\left(\frac{4g^{4}}{\kappa^{2}}-m^{2}\lambda^{2}\right)\left[\gamma + \ln\left(\frac{p^{2}}{\mu^{2}}\right)-i\pi\right]
-\frac{\bar{\theta}m\lambda^{2}}{8\pi}(1-\gamma + i\pi) 
\nonumber\\
&&+ \frac{\bar{\theta}g^4}{2\pi m\kappa^2}\left[\frac{1}{2}\ln(2\sin\varphi) + 2\cos\varphi\ln\left(\tan\left(\frac{\varphi}{2}\right)\right) -\frac{13}{3}-\frac{3}{4}\gamma + i\pi \right].
\end{eqnarray}
We see that at the critical point
\begin{equation}
\label{point}
\lambda=\pm \frac{2g^{2}}{m|\kappa|},
\end{equation}
dependence on the arbitrary mass scale $\mu$ disappears. As a result, the total
scattering amplitude becomes
\begin{eqnarray}
\label{ampl}
{\cal{A}}(\varphi)&=&-\frac{2ig^2}{m\kappa}\cot\varphi \mp \frac{2g^{2}}{m|\kappa|}
-\frac{\bar{\theta}g^{4}}{2\pi m\kappa^2}\left(\frac{1}{4}\gamma -\frac{16}{3}\right) 
\nonumber\\
&&+ \frac{\bar{\theta}g^4}{2\pi m\kappa^2}\left[\frac{1}{2}\ln(2\sin\varphi) + 2\cos\varphi\ln\left(\tan\left(\frac{\varphi}{2}\right)\right)\right].
\end{eqnarray}

The noncommutative AB scattering result by using coherent state
approach is  successfully obtained up to the one loop order. We have
shown that inclusion of a quartic self-interaction for the
scalar field allows to achieve a result possessing a smooth commutative limit. The choose of the lower or upper sign in (\ref{ampl}), corresponding to a attractive or repulsive quartic self-interaction.   

It is instructive to compare the scattering amplitude (\ref{ampl}) obtained within the coherent state approach with the results obtained for the scattering amplitude within the Moyal product approach \cite{Anacleto}.
The total scattering amplitude obtained within it reads as \cite{Anacleto}
\begin{eqnarray}
\label{loopm}
{\cal{A}}_{\mbox{1-loop}}(\varphi)
&=&-\frac{2ig^{2}}{m\kappa}\cot\varphi \mp \frac{2\sqrt{2}g^2}{m|\kappa|}
+\frac{ig^{4}}{2m\kappa ^{2}}+\frac{9g^{4}}{4\pi m\kappa ^{2}} 
+\frac{g^{4}}{2\pi m\kappa ^{2}}\ln[2\sin\varphi] 
\nonumber\\
&&+\frac{2\bar{\theta}g^{2}}{m\kappa}
+\frac{i\bar{\theta}g^{4}\sin\varphi}{\pi m\kappa ^{2}}
\ln\left[\tan\left( \frac{\varphi }{2}\right)\right].
\end{eqnarray}
Comparing the expressions (\ref{ampl}) and (\ref{loopm}) we arrive at the following conclusion.
Both amplitudes turn out to be free of any singularities at some critical relations of couplings, those are $\lambda=\pm\frac{2g^2}{m|\kappa|}$ within the coherent state approach and $\lambda=\pm \frac{\sqrt{2}g^2}{m|\kappa|}$ within the Moyal state approach. Thus we find that both approaches display the similar qualitative behaviour in the commutative limit. At the same time, we find that these amplitudes are different in some minor aspects. 
In the tree approximation and to first order in $\bar{\theta}$ our model does not present noncommutative correction whereas in \cite{Anacleto} it was shown a constant noncommutative contribution to
the two body scattering.
Furthermore, for small scattering angle $\varphi$, the noncommutative correction obtained in this paper
shows a $\ln\varphi$ dependence whereas in (\ref{loopm}) this dependence is $\varphi\ln\varphi$.
Also, we found that the third term in (\ref{ampl}) gives an constant noncommutative correction and this contribution does not present in (\ref{loopm}). 
However, these differences are quite natural being caused by use of different regularization and approximation schemes and by the different fashions of introduction of noncommutativity in two formulations of the theory, whereas the main feature, that is, absence of singularities at some certain relation of couplings, is observed in both theories.

\section{Summary and comments}
In this paper we have formulated the coherent state approach for the noncommutative AB scattering. 
In analogy with the noncommutative Moyal AB scattering \cite{Anacleto}  we observe that the total scattering
amplitude is UV finite. This happens due to arise of the natural cut-off  that appears in coherent state approach
while in the noncommutative Moyal we have UV and UV/IR singularities we need to implement the method of regularition in a conventional 
way introducing an ad hoc cut-off which has nothing to do with the parameter.

We show that noncommutative Aharonov-Bohm scattering amplitude obtained 
by using of the coherent state approach in the commutative limit 
agrees with the commutative result \cite{Lozano}. This is a solid argument for consistency of the 
method  used in this paper. Moreover, we have found that the principal physical conclusions, those are about absence of singularities to critical couplings, are valid both within the coherent state approach and the Moyal product approach. The problem of choice of the preferable approach between Moyal product formulation and the coherent states formulation needs special studies similarly to the problem of choice between Moyal product and Seiberg-Witten map \cite{Cha}.  

Let us also make some comments. First, we would note that the purely spacial character of the noncommutativity is essential for providing the unitarity of the theory, with the similar situation takes place in the relativistic Moyal-product based theories \cite{Gomis}. At the same time, the relativistic generalization of the coherent state method \cite{Spall} allows to provide unitarity without introducing privileged role of time.
Second, we can prove that the noncommutative corrections in the theory display exponential decay as distance grows. Indeed, the typical contributions arising within this approach, similar to (\ref{tri},\ref{box}), after integration in the internal momentum ${\bf{k}}$ but before of the expansion in power series in $\theta$, display the behaviour
\begin{eqnarray}
f(p)=\frac{e^{-\theta{\bf p}^2}}{({\bf p}^2+m^2)^a},
\end{eqnarray}
where ${\bf{p}}$ is the external momentum, and $a\geq 1$ is a some integer number. Let us transform this expression to the coordinate space. We get
\begin{eqnarray}
f(x)=\int\frac{d^2{\bf p}}{(2\pi)^2}\frac{e^{-\theta{\bf p}^2+i{\bf p}\cdot{\bf x}}}{({\bf p}^2+m^2)^a},
\end{eqnarray}
which after integration, for $m\to 0$, gives
\begin{eqnarray}
f(x)=\frac{1}{4\pi}(-\frac{d}{d{\bf x}^2})^a(\frac{1-e^{{-{\bf x}^2/\theta}}}{{\bf x}^2}).
\end{eqnarray}
We see that the $\theta$ dependent term representing itself as a noncommutative correction, first, vanishes at $\theta=0$, second, exponentially decreases as distance grows. So we conclude that the noncommutative corrections are highly suppressed with the distance.

\section{Acknowledgments}
The authors would like to thank M. Gomes for discussions. They are also grateful to D. Bazeia for calling their attention
to the coherent state method. This work was partially supported by PRONEX/CNPq/FAPESQ. 
M. A. A. was supported by CNPq/FAPESQ DCR program, project No. 350136/2005-0. A. Yu. P. was 
supported by CNPq/FAPESQ DCR program, project No. 350400/2005-9.

\newpage

\end{document}